\documentclass{ws-ijmpe}

\begin{document}
\graphicspath{{fig/}}

\markboth{Toshiki Maruyama}
{Non-Uniform Structure of Matter and the Equation of State}

\catchline{}{}{}{}{}

\title{NON-UNIFORM STRUCTURE OF MATTER AND THE EQUATION OF STATE}

\author{\footnotesize TOSHIKI MARUYAMA}

\address{
Advanced Science Research Center, Japan Atomic Energy Agency,\\
Shirakata Shirane 2-4, Tokai, Ibaraki 319-1195, Japan
\\
maruyama.toshiki@jaea.go.jp}

\author{SATOSHI CHIBA}

\address{
Advanced Science Research Center, Japan Atomic Energy Agency,\\
Shirakata Shirane 2-4, Tokai, Ibaraki 319-1195, Japan
}

\author{TOSHITAKA TATSUMI}

\address{Department of Physics, Kyoto University,
Kyoto 606-8502, Japan
}

\maketitle

\begin{history}
\received{(received date)}
\revised{(revised date)}
\end{history}

\begin{abstract}
We investigate the non-uniform structures and the equation of state (EOS) of
nuclear matter in the context of the first-order phase transitions (FOPT) such
as liquid-gas phase transition, kaon condensation, and hadron-quark phase transition.
During FOPT the mixed phases appear, where matter exhibits non-uniform
structures called ``Pasta'' structures due to the balance of the Coulomb repulsion
and the surface tension between two phases.
We treat these effects self-consistently, properly taking into account of the Poisson
equation and the Gibbs conditions. Consequently, they make the EOS of the mixed phase
closer to that of Maxwell construction due to the Debye screening.
This is a general feature of the mixed phase consisting of many species of
charged particles.

\end{abstract}


\section{Introduction}

Matter in stellar objects has a 
variety of densities and chemical components due to the presence of gravity.
At the surface of neutron stars,
there exists a region where the density is lower than the normal nuclear
density $\rho_0\simeq 0.16$fm$^{-3}$ over a couple of hundreds meters.
The pressure of such matter is retained by degenerate electrons,
while baryons are clusterized and have little contribution to the pressure.
Due to the gravity the pressure and the density increase
in the inner region
(in fact, the density at the center amounts to several times $\rho_0$ ).
Charge neutral matter consists of neutrons and the equal number of protons
and electrons under chemical equilibrium.
Since the kinetic energy of degenerate electrons is much higher than
that of baryons, the electron fraction (or the proton one) decreases
with increase of density and thus neutrons become the main component
and drip out of the clusters.
In this way baryons as well as electrons come to contribute to the pressure.
At a certain density, other components such as hyperons and 
mesons may emerge.
For example, negative kaon condensation, expected to be of a first-order
phase transition (FOPT), remarkably softens the equation of state of matter.
At even higher density, hadron-quark deconfinement transition may occur
and quarks in hadrons are liberated.
This phase transition is also considered to be of first-order.

\begin{figure}
\begin{minipage}{0.42\textwidth}
\includegraphics[width=0.90\textwidth]{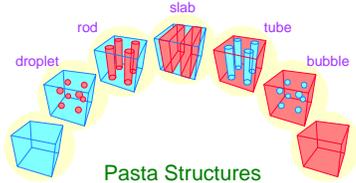}
\vspace{-5mm}
\end{minipage}\hspace{\fill}
\begin{minipage}{0.48\textwidth}
 \caption{Schematic picture of pasta structures.
 Phase transition from blue phase (left-bottom) to red phase (right-bottom) 
 is considered.
 }
\label{pasta}
\end{minipage}
\end{figure}

A FOPT brings about a thermodynamic instability of uniform matter to
have phase separation.
In other words, matter should have the nonuniform mixed phase (MP) around
the critical density.
Since nuclear matter consists of two chemically independent components, 
i.e.\ baryons and electrons,
the  equalities of both baryon and electron chemical potentials 
between two phases are required by the Gibbs conditions in the MP.
Therefore the EOS of the MP cannot be obtained simply by the
Maxwell construction, which is relevant only for single component.
Those components are electrically charged and non-uniformly distributed,
so that the local charge
neutrality is
no more held in the MP.
This point is important to the geometrical structure of the MP.
To minimize the surface energy plus the Coulomb energy,
matter is expected to form a structured mixed phase, i.e.\ a lattice of 
lumps of a phase with a geometrical symmetry embedded in the other phase.

At very low densities, nuclei in matter are expected to form the
Coulomb lattice embedded in the electron sea, that
minimizes the Coulomb interaction energy. 
With increase of density, ``nuclear pasta'' structures (see Fig.\ \ref{pasta}) 
emerge as a structured mixed phase\cite{Rav83} in the liquid-gas phase transition, 
where stable nuclear shape may change from droplet to rod, slab, tube, and to bubble. 
Pasta nuclei are eventually dissolved into uniform matter at a certain nucleon 
density below the saturation density $\rho_0$. 
The name ``pasta'' comes from rod and slab structures 
figuratively spoken as ``spaghetti'' and ``lasagna''.
Such low-density nuclear matter exists in the collapsing stage of
supernovae and in the crust of neutron stars.
Supernova matter is relevant to liquid-gas transition
of non-beta-equilibrium nuclear matter with a fixed proton fraction
and the low-density neutron star matter is relevant to 
neutron-drip transition of beta-equilibrium nuclear matter.
The structured mixed phase is also expected in the phase transitions at
higher density, 
like kaon condensation and 
hadron-quark phase transition.
In these cases, the charge screening effect may be pronounced 
because the local charge density can be high.

Our purpose here is 
to investigate low-density nuclear pasta structure, kaonic pasta structure,
and hadron-quark pasta structure self-consistently within  
the mean-field approximation. 
In particular, 
we figure out how the Coulomb screening and the surface tension
affect the property of the MP.

\section{Low Density Nuclear Matter}

First we investigate the property of nuclear matter
at low density.
Exploiting the idea of the density functional theory within 
the relativistic mean field (RMF) model, we can formulate equations of motion 
to study non-uniform nuclear matter numerically, cf.\ Refs.\ \refcite{drez,parr}.
The RMF model with fields of mesons and baryons introduced
in a Lorentz-invariant way is not only relatively simple for numerical calculations,
but also sufficiently
realistic  to reproduce bulk properties of finite nuclei
as well as the saturation properties of nuclear matter.\cite{maru05,marurev}
In our framework, the Coulomb interaction is properly included in the 
equations of motion for nucleons and electrons and for meson mean fields,
and we solve the Poisson equation for the Coulomb potential $V$
self-consistently with those equations.
Thus the baryon and electron density profiles, as well as the meson
mean fields, are determined in a fully
consistent way with the Coulomb interaction.

\subsection{Relativistic mean field model}
\subsubsection{Thermodynamic potential and the equations of motion}

We start with the thermodynamic potential for
the system of neutrons, protons, electrons and mesons 
with temperature $T=0$,
\begin{equation}\label{Omega-tot}
\Omega = \Omega_N+\Omega_M +\Omega_e.
\end{equation}
The first term is the contribution of nucleons 
with the local-density approximation,
%
\begin{eqnarray}
\Omega_N  &=&
  \sum_{a=p,n}
  \int  d^3r
  \left[
  \int_0^{k_{{\rm F},a}}
  { d^3k \over 4\pi^3}
  \sqrt{{m_N^*}^2+k^2}-\rho_a\nu_a
  \right],
\label{OmegaN}\\
\nu_n({\bf r})&=&\mu_n-g_{\omega N}\omega_0({\bf r})+g_{\rho N}R_0({\bf r}),\\
\nu_p({\bf r})&=&\mu_p+
{V({\bf r})}-g_{\omega N}\omega_0({\bf r})-g_{\rho N}R_0({\bf r}),
\end{eqnarray}
with the local Fermi momenta
$k_{{\rm F},a}({\bf r})$ ($a=n,p$), the effective nucleon mass $m_N^*({\bf r})=m_N-g_{\sigma N}\sigma({\bf r})$,
%
%
the chemical potentials $\mu_a$ ($a=n,p$),
and coupling constants 
$g_{\sigma N}$, $g_{\omega N}$ and  $g_{\rho N}$.

The second term  in  (\ref{Omega-tot}) includes
the scalar ($\sigma$) and vector ($\omega_0, R_0$) mean fields,
%
\begin{equation}
\Omega_M = \int   d^3r \Biggl[
  {(\nabla\sigma)^2 + m_\sigma^2\sigma^2 \over2} + U(\sigma) 
  -{(\nabla\omega_0)^2 + m_\omega^2\omega_0^2 \over2}
  -{(\nabla R_0)^2 + m_\rho^2R_0^2\over2}  \Biggr]  ,
\label{OmegaM}
\end{equation}
where $m_\sigma$, $m_\omega$ and $m_\rho$ are the field masses, and
$U(\sigma)={1\over3}bm_N(g_{\sigma N}\sigma)^3+{1\over4} c(g_{\sigma N}\sigma)^4$
is the nonlinear potential for the scalar field.

The third term in (\ref{Omega-tot}) contains the contribution of
the Coulomb field  (described by the potential $V({\bf r})$)
and the contribution of relativistic electrons,
\begin{equation}
\Omega_e = \int d^3r \left[
-{1\over8\pi e^2}(\nabla {V})^2-{(\mu_e-{V})^4\over12\pi^2}
\right],
\end{equation}
where $\mu_e$ is the electron chemical potential.


Equations of motion for the fields 
are obtained
from the variational principle, 
%
\begin{eqnarray}
\nabla^2\sigma({\bf r}) &=& m_\sigma^2\sigma({\bf r}) +{dU\over d\sigma}
-g_{\sigma N}(\rho_n^{s}({\bf r}) +\rho_p^{s}({\bf r}))
,\label{EOMsigma}\\
\nabla^2\omega_0({\bf r}) &=& m_\omega^2\omega_0({\bf r}) 
-g_{\omega N} (\rho_p({\bf r})+\rho_n({\bf r}))
,\label{EOMomega}\\
\nabla^2R_0({\bf r}) &=& m_\rho^2R_0({\bf r}) -g_{\rho N}
(\rho_p({\bf r})-\rho_n({\bf r}))
,\label{EOMrho}\\
\nabla^2V({\bf r}) &=& 4\pi e^2\rho_{\rm ch}({\bf r}),
\\
\mu_e&=&\left({3\pi^2}\rho_e({\bf r})\right)^{1/3}+V({\bf r}),\label{EOMmuE}\\
\mu_n\;=\;\mu_B&=&\sqrt{k_{{\rm F},n}({\bf r})^2+{m_N^*({\bf r})}^2}
+g_{\omega N}\omega_0({\bf r})
-g_{\rho N}R_0({\bf r}) \label{EOMmuN},\\
\mu_p\;=\;\mu_B-\mu_e&=&\sqrt{k_{{\rm F},p}({\bf r})^2+{m_N^*({\bf r})}^2}
+g_{\omega N}\omega_0({\bf r})
+g_{\rho N}R_0({\bf r})-V({\bf r}).
\label{EOMmuNp}
\end{eqnarray}
with 
the proton and neutron scalar densities 
and the charge density 
\begin{eqnarray}
\rho_a^{s}({\bf r})&=& 
\int_0^{k_{{\rm F},a}({\bf r})} \frac{d^3k}{4\pi^3} 
\frac{m_N^*({\bf r)}}{\sqrt{m_N^*({\bf r})^2+k^2}}, \quad (a=p,n),
\\
\rho_{\rm ch}({\bf r})&=& \rho_p({\bf r})-\rho_e({\bf r}).
\label{eq:poisson}
\end{eqnarray}

To solve the above coupled equations numerically,
the whole space is divided into equivalent Wigner-Seitz cells with a radius $R_{\rm W}$.
The geometrical shape of the cell changes as follows:
sphere in three-dimensional (3D) calculation, cylinder in 2D and slab in 1D, respectively.
Each cell is globally charge-neutral and all physical quantities
in the cell are smoothly connected to those of the next cell
with zero gradients at the boundary.
Every point inside the cell is represented by the grid
points (number of grid points $N_{\rm grid}\approx 100$) and
differential equations for fields are solved by the relaxation method
for a given baryon-number density under constraints of the global charge neutrality.
Details of the numerical procedure are explained in Refs.\ \refcite{maru05,marurev}.


Parameters of the RMF model are chosen to reproduce saturation properties
of symmetric nuclear matter:
the minimum energy per nucleon $-16.3$ MeV at $\rho =\rho_0 \equiv 0.153$ fm$^{-3}$,
the incompressibility $K(\rho_0) =240$ MeV, the effective nucleon  mass
$m_N^{*}(\rho_0)=0.78 m_N$; $m_N =938$ MeV, and the isospin-asymmetry
coefficient $a_{\rm sym}=32.5$ MeV.
Coupling constants and meson masses used in our calculation are listed in Table 1.

\begin{table}[t]
\begin{center}
\caption{
Parameter set used in RMF calculation in Secs.\ 2 and 3.
The kaon optical potential $U_K$ is
defined by $U_K = g_{\sigma K}\sigma+g_{\omega K}\omega_0$.
}
\par\medskip
{\tablefont
\begin{tabular}{@{}cccccccccccc@{}}
\toprule
$g_{\sigma N}$ &
$g_{\omega N}$ &
$g_{\rho N}$ &
$b$ &
$c$ &
$m_\sigma$ &
$m_\omega$ 
\\
\colrule
6.3935 &
8.7207 &
4.2696 &
0.008659 &
0.002421 &
 400 MeV &
 783 MeV
\\
\botrule
\end{tabular}
\par\ \par
\begin{tabular}{@{}cccccc@{}}
\toprule
$m_\rho$ &
 $f_K (\approx f_\pi)$ &
 $m_K$ &
 $g_{\omega K}$ &
 $g_{\rho K}$ &
 $ U_K(\rho_0)$ 
\\ [1mm]
\colrule
 769 MeV &
 93 MeV  &
 494 MeV &
$g_{\omega N}/3$ &
$g_{\rho N}$ &
$-130$ MeV \\
\botrule
\end{tabular}
}
\end{center}
\end{table}

For the study of non-uniform nuclear matter, the ability to reproduce
the bulk properties of finite nuclei should be essential.
We have checked how it works to describe finite nuclei
and
we have checked that the bulk properties of finite nuclei 
(density, binding energy, and proton to baryon number ratio) 
are satisfactorily reproduced for our present purpose.\cite{maru05,marurev}

\subsubsection{Control of surface tension}

Note that in our framework we must use a sigma mass 
$m_{\sigma}=400$ MeV,\cite{centelles93} a slightly smaller value 
than that one usually  uses, to get an appropriate fit. 
If we used a popular value $m_\sigma\approx 500$ MeV, finite nuclei would be
over-bound by about 3 MeV/$A$. 
The actual value of the sigma
mass (as well as the omega and rho masses) has little relevance
for the case of infinite nucleon matter, since it enters the
thermodynamic potential only in the combination
$\widetilde{C}_{\sigma}=g_{\sigma N}/m_{\sigma}$. 
However meson masses  are important characteristics of finite nuclei and of
other non-uniform nucleon systems, like those in pasta. 
The effective meson mass characterizes the typical scale for the
spatial change of the meson field and consequently it affects the
value of the effective surface tension.

If we artificially multiply meson masses
$m_{\sigma}$, $m_{\omega}$ and $m_{\rho}$  by a factor $c_M$,
e.g.\ $c_M =1$ (realistic case), 2.5 and 5.0,
the surface tension changes.
By the use of heavy meson masses, the binding energy
of finite nuclei (for finite $A$)
approaches to that of nuclear matter. 
This shows that the surface tension is reduced
with increase of the meson masses, cf.\ Ref.\ \refcite{MS96}.
Notice that this statement is correct only if we
fix the ratio $g_{\phi N}^2 /m_\phi^2$.
Using the above modified meson masses, we explore the effects of
the surface tension later.

\subsection{Nucleon matter with fixed proton fraction}

\begin{figure*}
\begin{minipage}{0.55\textwidth}
\includegraphics[width=.49\textwidth]{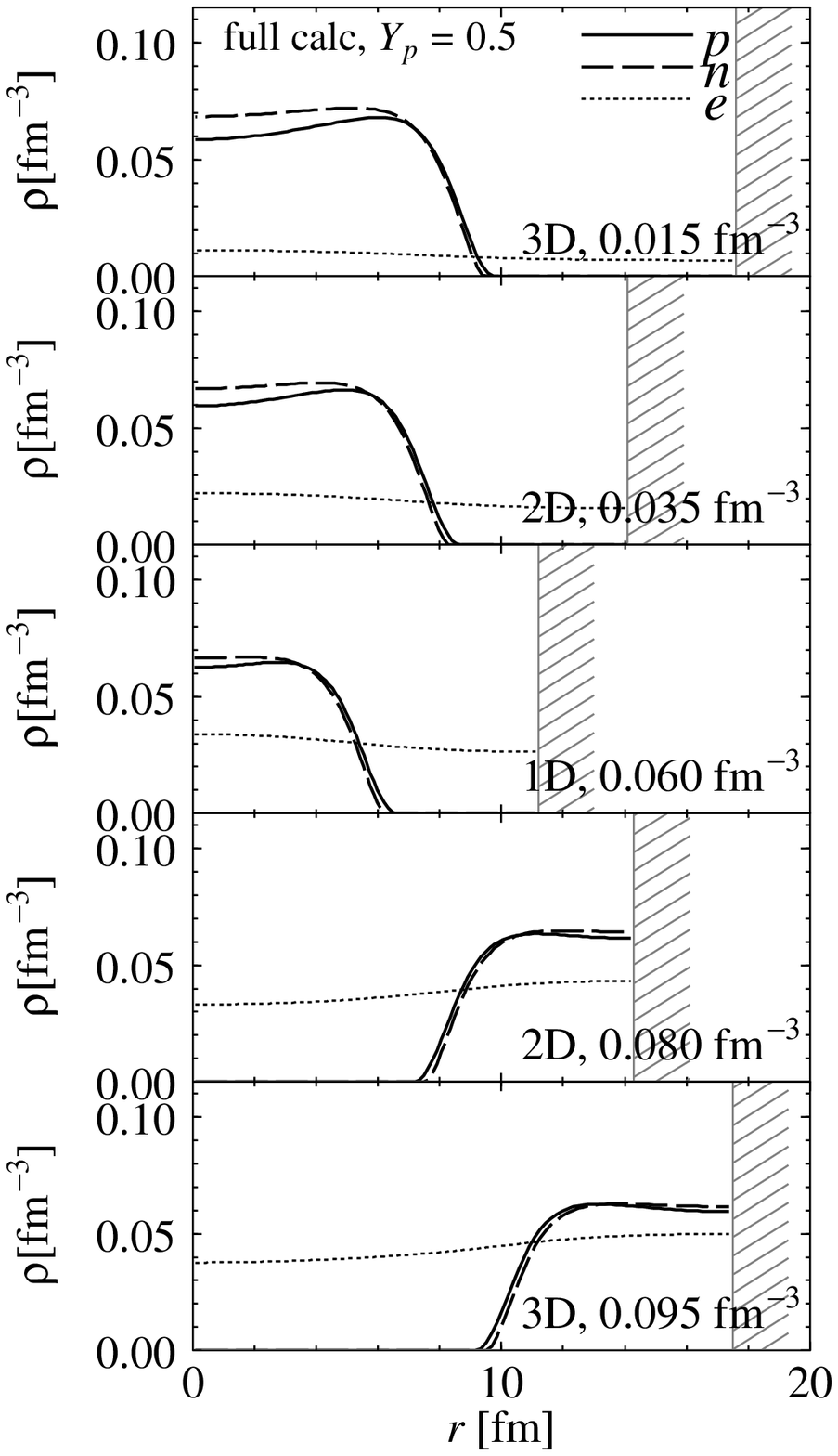}
\includegraphics[width=.49\textwidth]{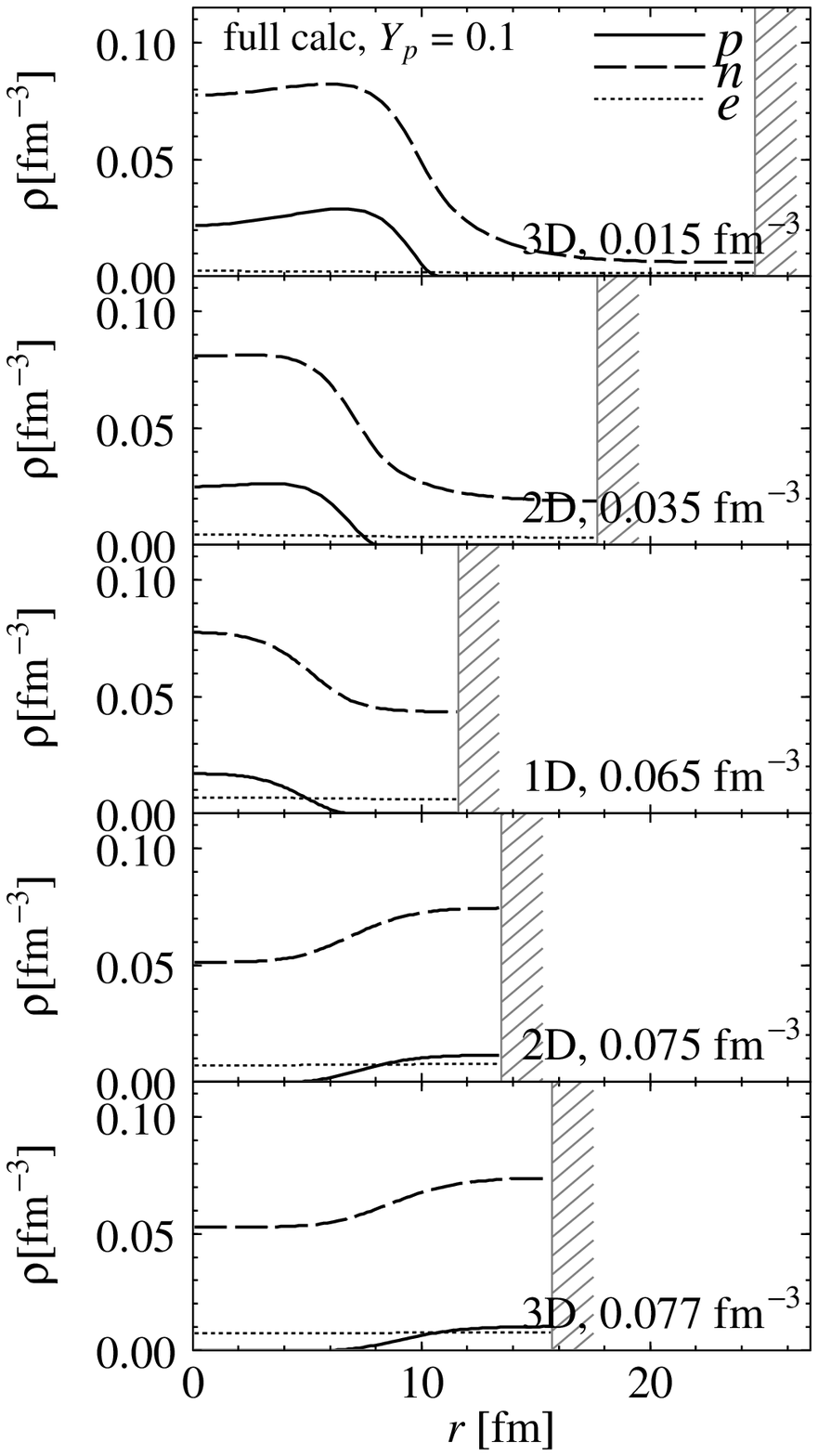}
\end{minipage}\hspace{\fill}
\begin{minipage}{0.38\textwidth}
\caption{
Examples of the density profiles in the cell for symmetric nuclear
 matter with $Y_p$=0.5 (left) and for asymmetric matter
 with 
 $Y_p=0.1$ (right).
}
\label{proffixfull}
\end{minipage}
\end{figure*}

First, we concentrate on the behavior of 
nucleon matter at fixed values of the proton fraction $Y_p$.
Particularly, we explore $Y_p=0.1$
and 0.5.
The case 
$Y_p=0.5$ should be relevant for supernova matter
and for newly born neutron stars. 
Figure \ref{proffixfull} shows
some typical density profiles inside the Wigner-Seitz cells. 
The geometrical dimension of the cell  is denoted as ``3D''
(three-dimensional sphere), etc. 
The horizontal axis in each panel
denotes the radial distance from the center of the cell. 
The cell boundary is indicated by the hatch.
 From the top to the bottom the configuration corresponds to
droplet (3D), rod (2D), slab (1D), tube (2D), and bubble (3D).
The nuclear ``pasta'' structures are clearly manifested.
For the lowest $Y_p$ case ($Y_p =0.1$), the neutron density is finite at any point:
the space is filled by dripped neutrons.
The value of $Y_p$ above which neutrons drip
is around 0.26 in our 3D calculation, for example.
For a higher  $Y_p$, the neutron density
drops to zero outside the nucleus. 
The proton number density always drops to zero outside the nucleus.
We can see that the charge screening effects are pronounced. 
Due to the spatial rearrangement of electrons the electron
density profile becomes no more uniform. 
This non-uniformity of the electron distribution is
more pronounced for a higher $Y_p$ and a higher density. 
Protons repel each other.
Thereby the proton density profile substantially deviates from the step-function. 
The proton number is enhanced 
near the surface of the lump.

\begin{figure*}
\begin{minipage}{0.58\textwidth}
\includegraphics[width=.49\textwidth]{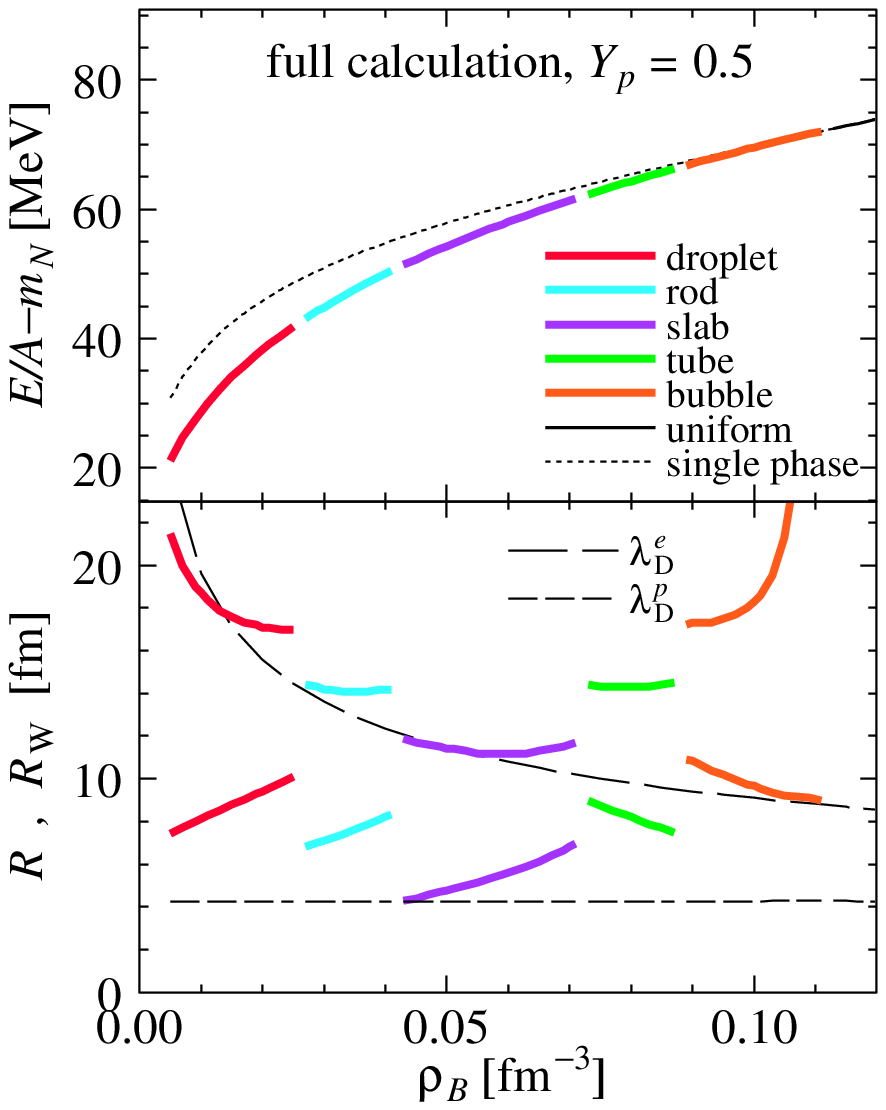}
\includegraphics[width=.49\textwidth]{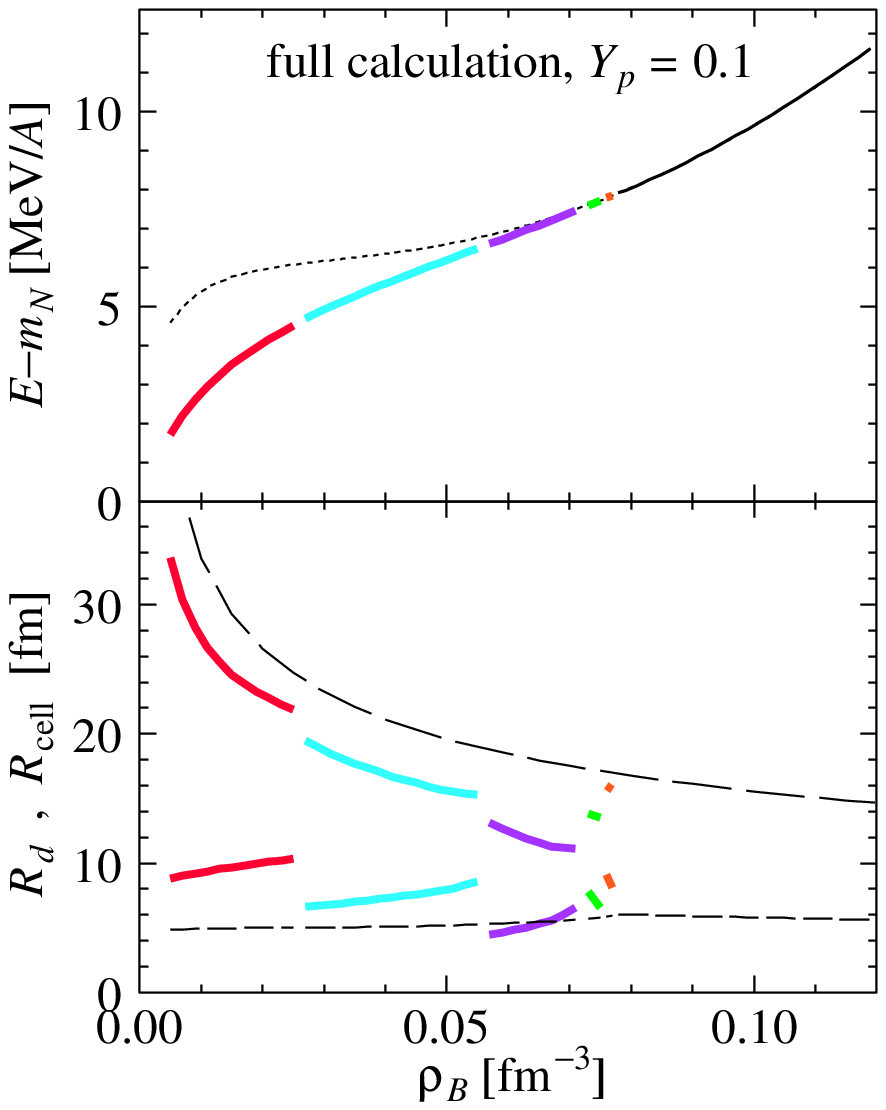}
\end{minipage}\hspace{\fill}
\begin{minipage}{0.38\textwidth}
\caption{
Binding energy per nucleon (upper) and the cell and lump sizes (lower) 
for symmetric nuclear matter with $Y_p$=0.5 (left), and for asymmetric matter with
$Y_p=0.1$ (right).
}
\label{eosfixfull}
\end{minipage}
\end{figure*}

The equation of state (EOS) for the sequence of geometric structures is
shown in Fig.~\ref{eosfixfull} (upper panels)
as a function of the averaged baryon-number density.
Note that the energy $E/A-m_N$ also includes the
kinetic energy of electrons, which makes the total pressure positive. 
The lowest-energy configurations are selected among
various geometrical structures. 
The most favorable configuration
changes from the droplet to rod, slab, tube, bubble, and to the
uniform one (the dotted thin curve) with an increase of density.
The appearance of non-uniform structures in matter results in a
softening of EOS: the energy per baryon
gets lower up to about 15 MeV$/A$ compared to uniform matter.

The lower panels in Fig.~\ref{eosfixfull} are the cell radii
$R_{\rm W}$ and the lump radii $R$ versus averaged baryon number density. 
Dashed curves show the Debye screening lengths of
electrons and protons calculated as
$ 
\lambda^{e(p)}_{\rm D} = \left(4\pi e^2{d\rho_{e(p)}^{\rm av}\over d\mu_{e(p)}}\right)^{-1/2}.
$ 
In all cases, except for bubbles (at $Y_p=0.5$),
$R$ are
smaller than ${\lambda}_{\rm D}^{e}$.
This means that the Debye screening effect of electrons inside 
these structures should not be pronounced.

Let us mention briefly about the neutron star matter 
in beta equilibrium. 
The apparently different feature in this case is
that only the droplet configuration appears as a non-uniform structure. 
It should be noticed, however, that the presence or
absence of the concrete pasta structure sensitively depends on the
choice of the effective interaction.
The droplet structure is
quite similar to the case of the fixed proton mixing 
ratio $Y_p=0.1$ considered above. 
The screening effect is very small due to the small proton ratio.\cite{maru05,marurev}

\section{Kaon Condensation at High Densities}

Next let us explore high-density nuclear matter in beta equilibrium, 
which is expected in the inner core of neutron stars.

Kaons are Nambu-Goldstone bosons accompanying the spontaneous
breaking of chiral $SU(3)\times SU(3)$ symmetry 
and the lightest mesons with strangeness. 
Their effective energy is much reduced by the kaon-nucleon interaction 
in nuclear medium, which is dictated by chiral symmetry.
For low-energy kaons the $s$-wave interaction is dominant and attractive in
the $I=1$ channel, so that negatively charged kaons appear in the neutron-rich
matter once the process $n\rightarrow p+K^-$ becomes energetically allowed.
Since kaons are bosons, they cause the Bose-Einstein condensation 
at zero momentum.\cite{kn86}
The single-particle energy of kaons is given in a model-independent way: 
$ 
\epsilon_\pm({\bf p})=\sqrt{|{\bf p}|^2+m_K^{*2}+((\rho_n+2\rho_p)/4f^2)^2}\pm (\rho_n+2\rho_p)/4f^2,
$ 
where $m_K^*$ is the effective mass of kaons,
$m_K^{*2}=m_K^{2}-\Sigma_{KN}(\rho_n^{s}+\rho_p^{s})/f^2$, with the $KN$ sigma term,
$\Sigma_{KN}$, and the meson decay constant, $f\equiv f_K\sim f_\pi$.
The threshold condition then reads\cite{mut}
$ 
\mu_K=\epsilon_-({\bf p}=0)=\mu_n-\mu_p=\mu_e,
$ 
which means the kaon distribution function diverges at ${\bf p}=0$. 

\subsection{RMF treatment of nuclear matter with kaon condensation}
We explore high-density nuclear matter with kaon condensation by means of
RMF model as in low-density matter.
Using the same model we can discuss the non-uniform structure of nuclear matter
both at low- and high-density regime in an unified way.
To incorporate kaons into our RMF calculation,
the thermodynamic potential of Eq.~(\ref{Omega-tot}) is modified as
\begin{eqnarray}
\Omega &=& \Omega_N+\Omega_M +\Omega_e +\Omega_K,\\
\Omega_K&=&\!\!\!\int d^3r\left\{
  {f_K^2(\nabla\theta)^2\over2}
  -{f_K^2\theta^2\over2}\left[-{m_K^*}^2+(\mu_K-V+g_{\omega K}\omega_0
  g_{\rho K}R_0)^2 \right]
  \right\},
\end{eqnarray}
where $m_K^*=m_K-g_{\sigma K}\sigma$, $\mu_K=\mu_e$,  
and the kaon field $K=f_K\theta/\sqrt{2}$ ($f_K$ is the kaon decay constant). 
We, hereafter, neglect a rather unimportant term $\propto \sigma^2\theta^2$.%
The equations of motion are similar to Eqs.~(\ref{EOMsigma}) - (\ref{EOMmuNp}) 
given for the low-density case (Sec.\ 2).
See Refs.\ \refcite{maruKaon,marurev} for details.
Additional parameters 
are presented in Table 1.

\subsection{Kaonic pasta structures}

\begin{figure*}[b]
\begin{minipage}{0.48\textwidth}
  \includegraphics[width=.80\textwidth]{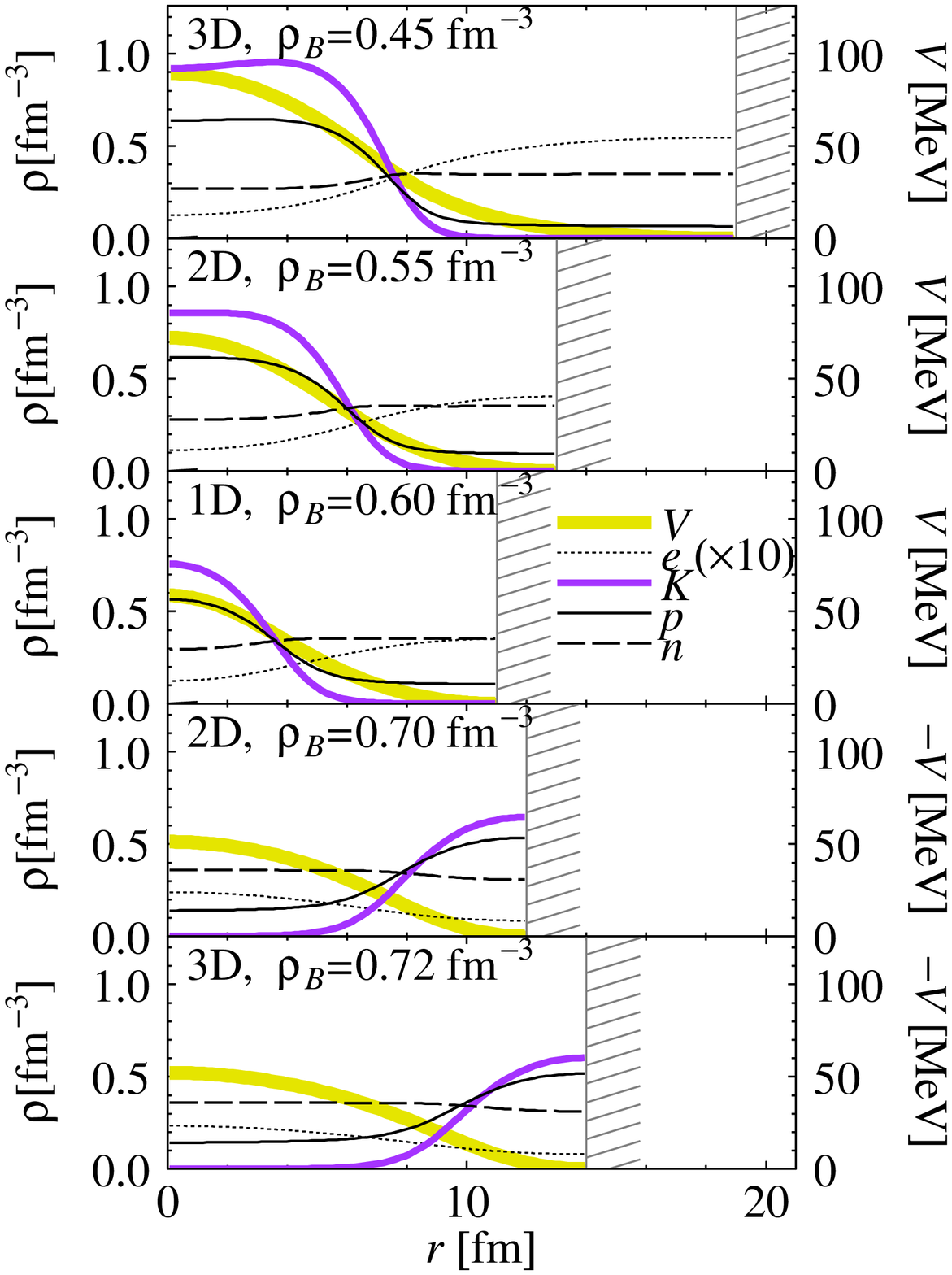}
  \caption{
   Density profiles of kaonic pasta structures.
  Here the density does not mean charge-density but number-density
  of particles.
  The kaon optical potential $U_K$ at the nuclear saturation
  density is set to be $-130$ MeV.
}\label{figProfK}
\end{minipage}
\hspace{\fill}
\begin{minipage}{0.48\textwidth}
  \includegraphics[width=.78\textwidth]{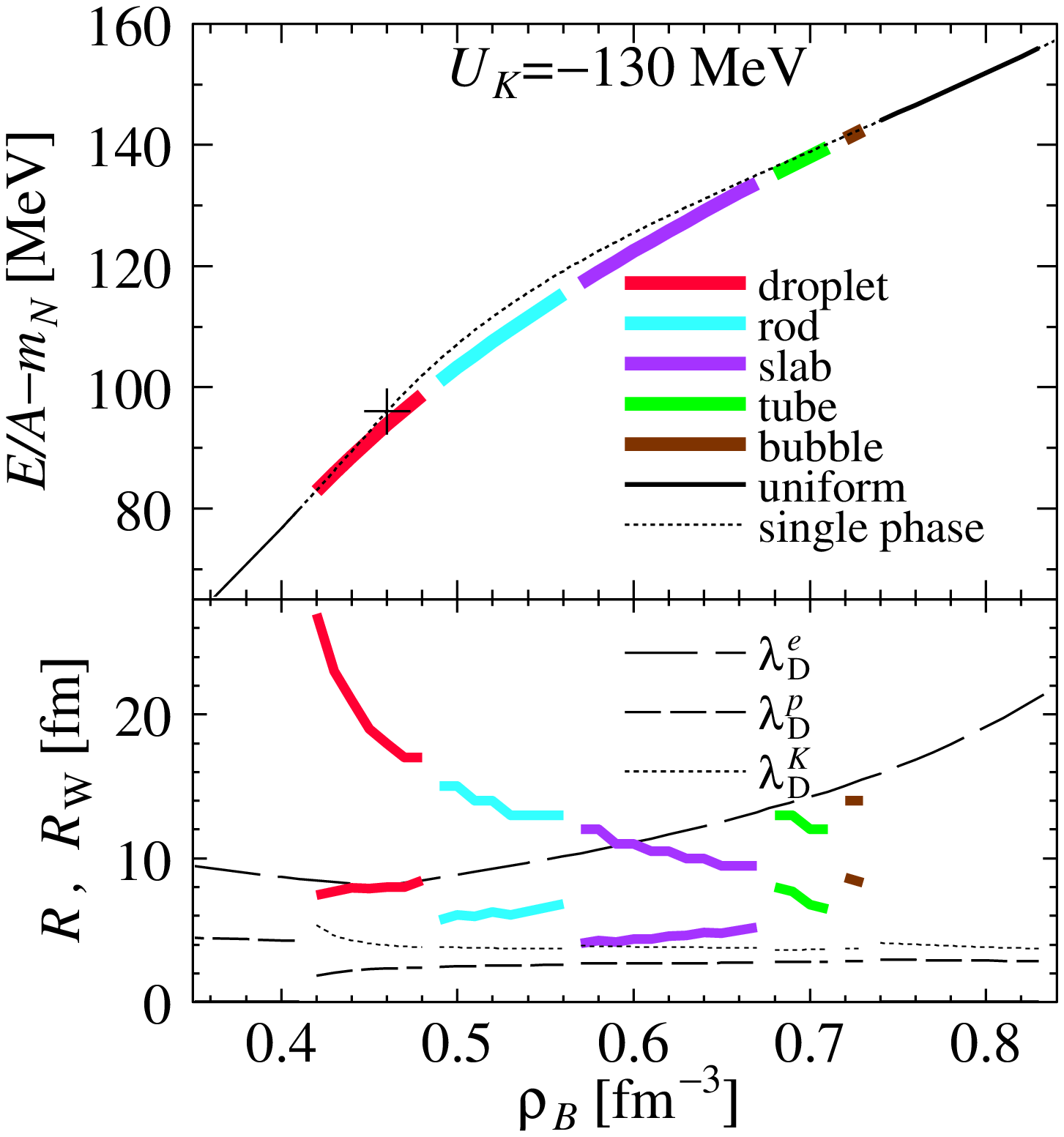}
  \caption{
  Upper: binding energy per nucleon of the
  nuclear matter in beta equilibrium.
  The dotted line below the cross shows uniform normal nuclear matter
  and above the cross, uniform kaonic matter.
  Lower: lump size $R$ (thick curves below)
  and cell size $R_{\rm W}$ (thick curves above).
  Compared are the Debye screening lengths of electron, proton and kaon.
  The Debye screening lengths are calculated using the
  explicit dependence of $\mu_a$ on $\rho_a$. 
}\label{figKEOS}
\end{minipage}
\end{figure*}

Figure \ref{figProfK} displays typical density profiles.
The neutron distribution proves to be rather flat.
The proton distribution on the other hand is strongly
correlated with the kaon distribution,
which means that the Coulomb interaction is crucial.

In the upper panel of Fig.\ \ref{figKEOS} we depict the energy per
nucleon as a function of baryon number density.
The dotted line indicates the case of single phase
(if one assumes the absence of the mixed phase).
In this case uniform matter consists of
normal nuclear matter below the critical density
and kaonic matter above the critical density.
The cross  on the dotted line ($\rho_B \simeq 0.46$~fm$^{-3}$)
shows the critical density, i.e.\ the point where kaons begin
to condensate in the case of single phase.

The lower panel of Fig.\ \ref{figKEOS} shows the sizes of the lump $R$ and 
the cell $R_{\rm W}$.
The dashed lines and the dotted line in the lower panel of Fig.~\ref{figKEOS}
show partial contributions to the Debye screening lengths of
the electron, proton and kaon,
$\lambda_{\rm D}^{e}$, $\lambda_{\rm D}^{p}$,
and $\lambda_{\rm D}^{K}$, respectively.
We see that in most cases $\lambda_{\rm D}^{e}$ is less than the cell size
$R_{\rm W}$ but is larger than the lump size $R$.
The proton Debye length $\lambda_{\rm D}^{p}$
and the kaon Debye length $\lambda_{\rm D}^{K}$, on the other hand,
are always shorter than $R_{\rm W}$ and $R$.

\section{Charge Screening and Surface Effects on the Pasta Structures}


\begin{figure}[b]
\includegraphics[width=.32\textwidth]{eos-full05-5BW-.ps}
\includegraphics[width=.32\textwidth]{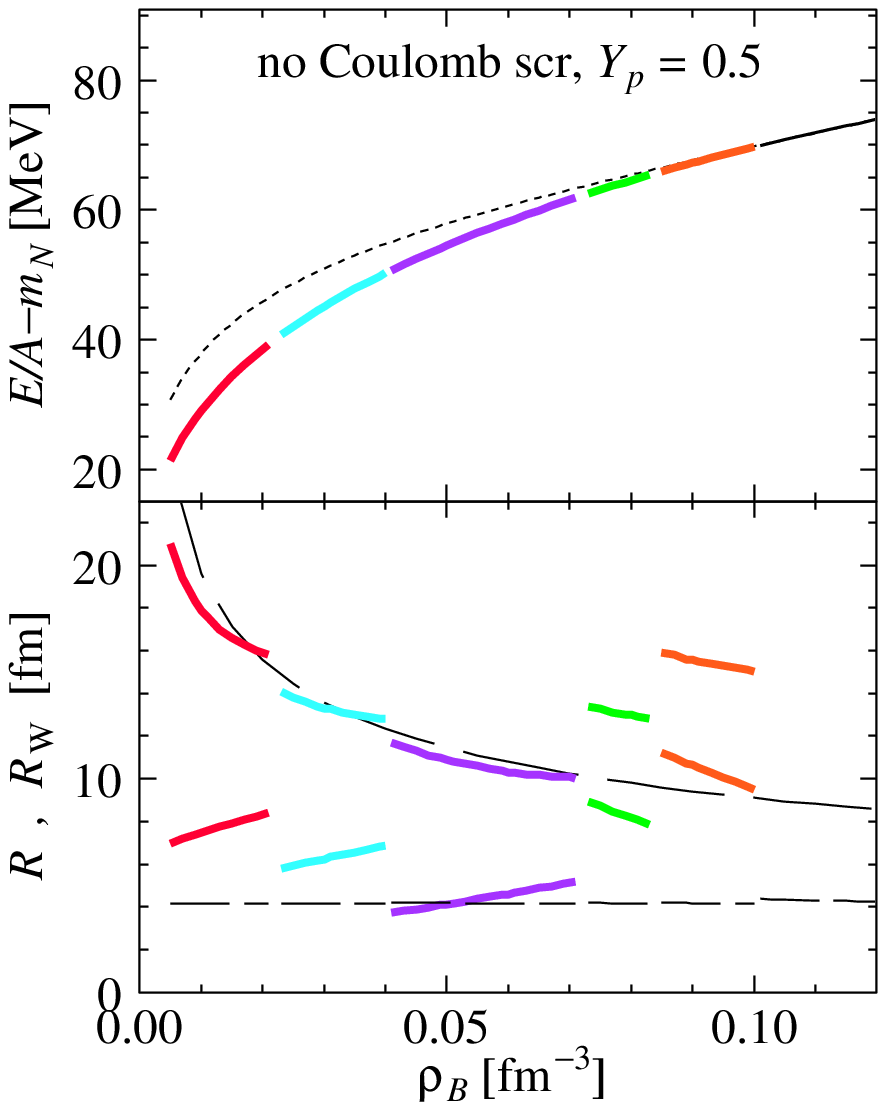}
\includegraphics[width=.32\textwidth]{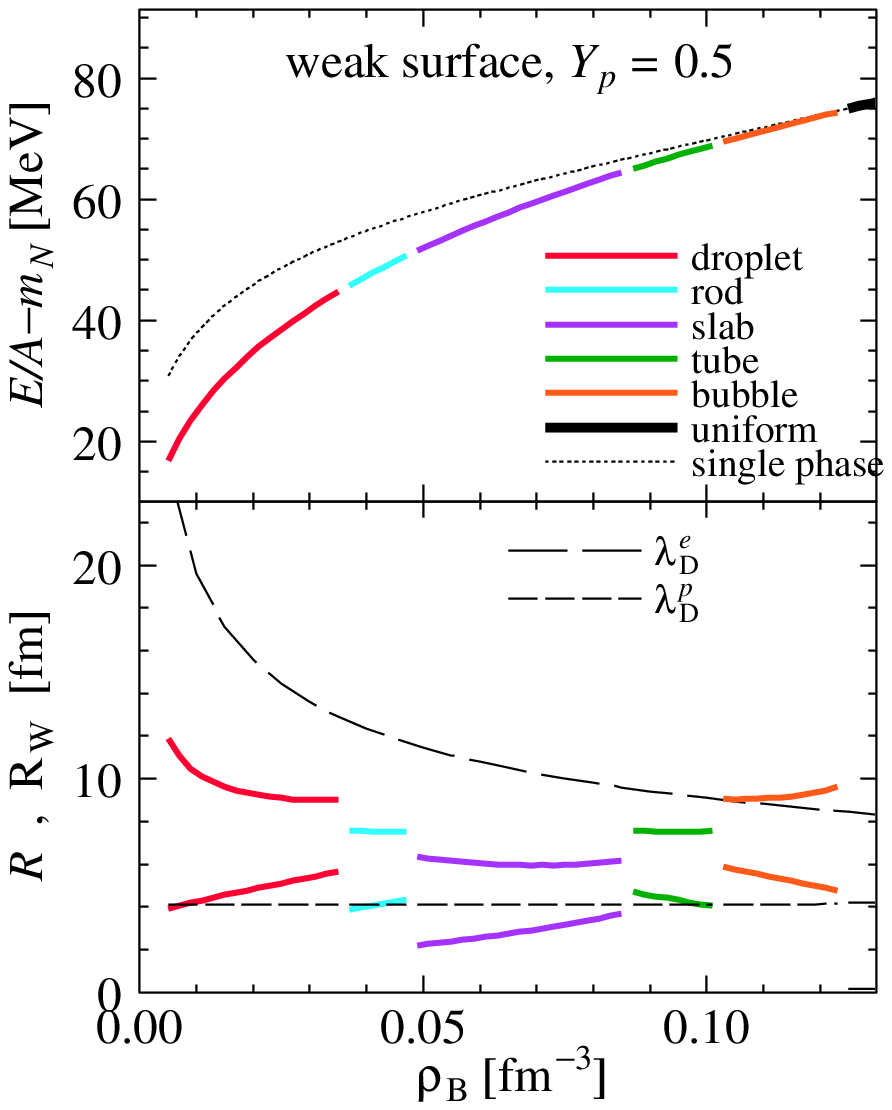}
\caption{
Upper: binding energy per nucleon of nuclear matter.
Lower: lump size $R$ and cell size $R_{\rm W}$.
The proton mixing ratio is $Y_p$=0.5 for all cases.
 From the left: ``full'', ``no Coulomb screening'' and ``weak surface'' calculations.
 In ``no Coulomb screening'' calculation, the electric potential is discarded 
 when determining the density profile
 and then added to evaluate the energy.
}
\label{eoscompare}
\bigskip
%
  \includegraphics[width=.32\textwidth]{Eos2WC130LK-.ps}
  \includegraphics[width=.32\textwidth]{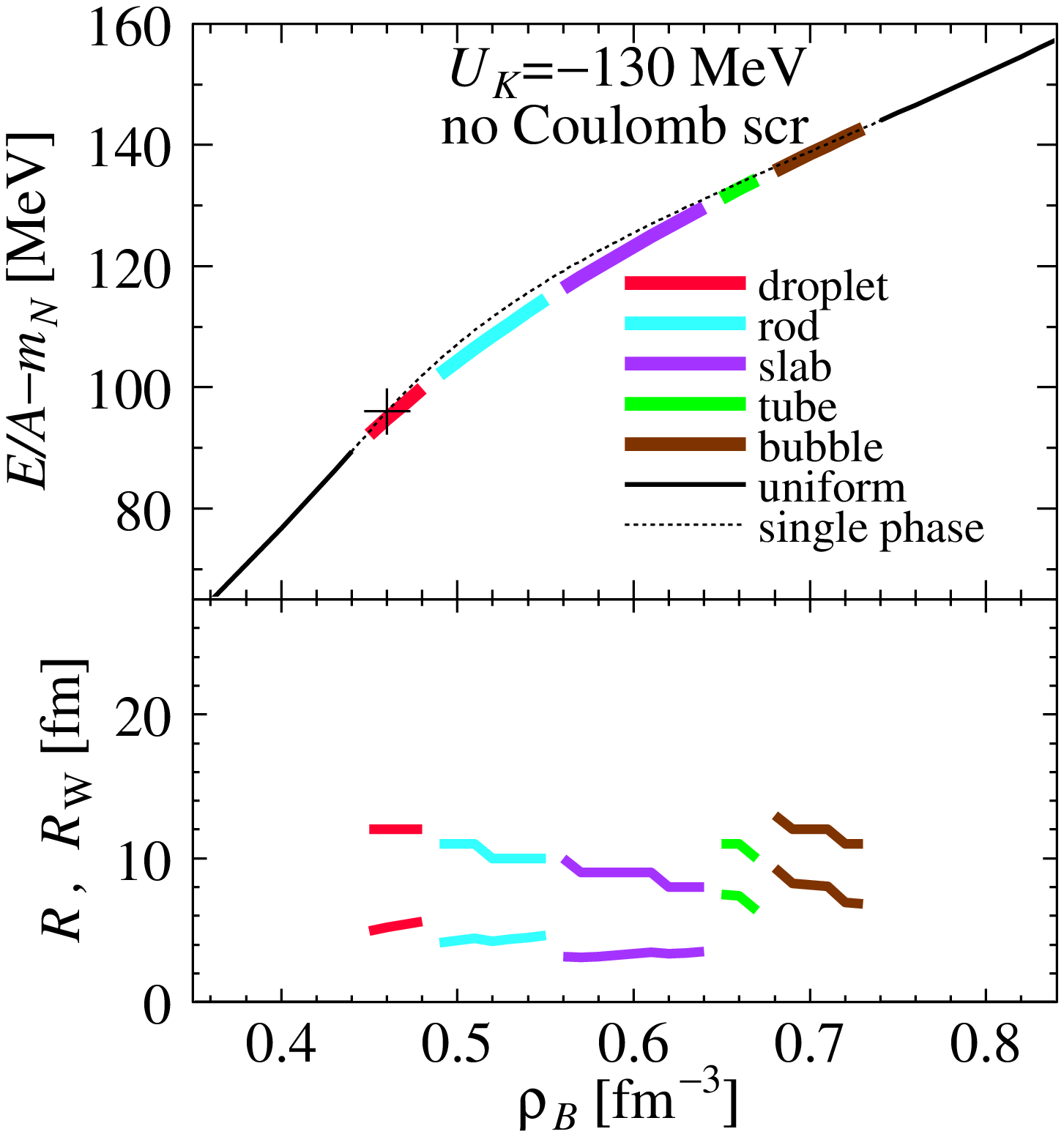}
  \includegraphics[width=.32\textwidth]{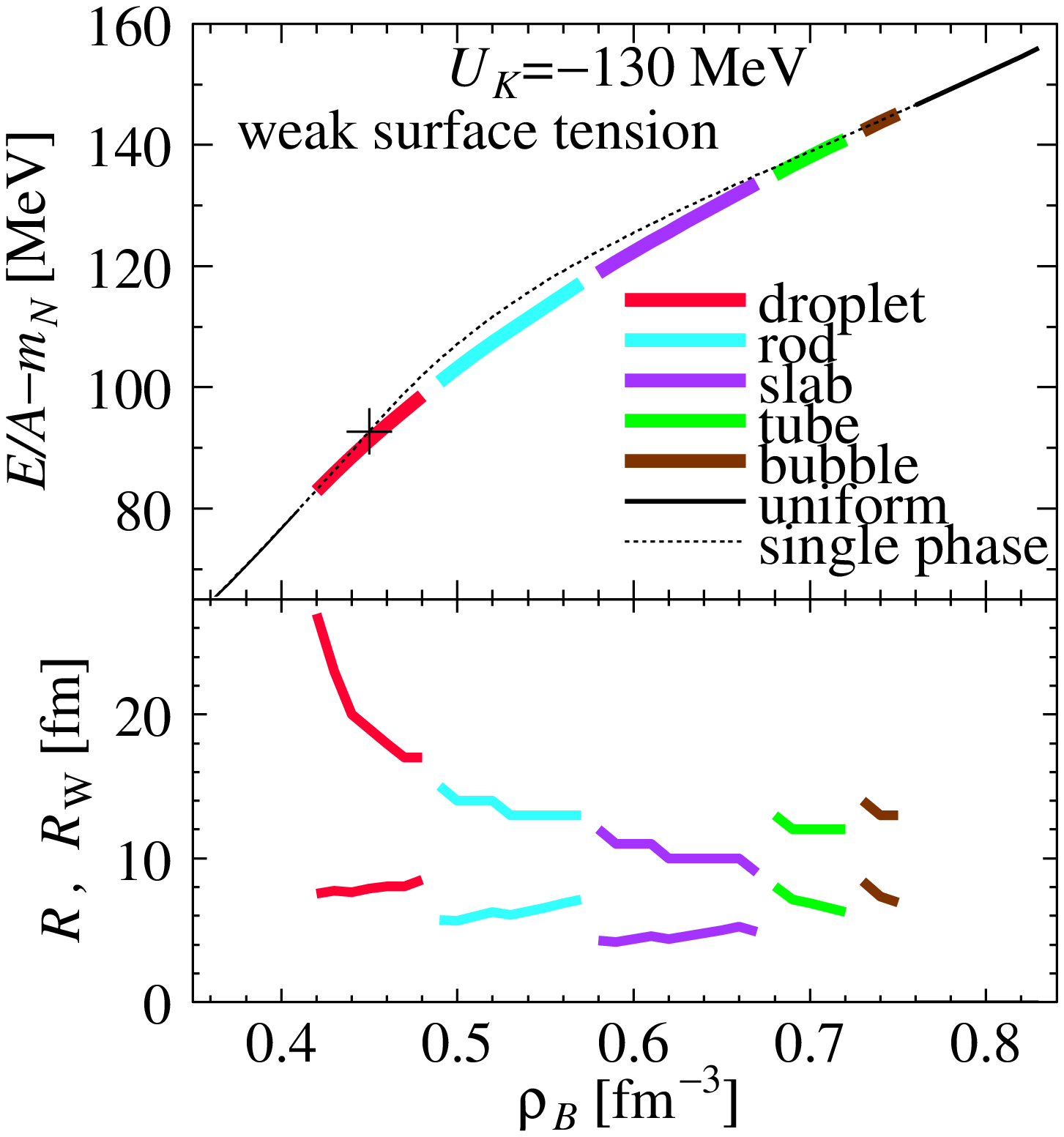}
\caption{
The same as Fig.\ \ref{eoscompare} for high-density nuclear matter
in beta equilibrium.
}\label{figKEOScompare}
\end{figure}

To demonstrate the charge screening effects we compare
results of the full calculation with
those given by a perturbative treatment of the Coulomb interaction often 
used in the literature, which we refer to as ``no Coulomb screening'' calculation.
In this calculation the electric potential is discarded in equations of motion
which determine the density profiles.
The Coulomb energy is then added to the total energy by using the
charge density profile thus determined to find the optimal value
with respect to the cell size $R_{\rm W}$.%

First, we discuss the case of low-density symmetric nuclear matter.
In the left and the central panels of Fig.~\ref{eoscompare},
compared are different treatments of the Coulomb interaction.
The EOS (upper panels) as a whole shows
almost no dependence on the treatments of the Coulomb interaction.
However, sizes of the cell and the lump (lower panels),
especially for tube and bubbles, are different.
The other effect
is a  difference in the density range for
each pasta structure.
The ``full'' treatment of the Coulomb
interaction slightly increases the density region of the nuclear pasta.

We show the same comparison for the kaonic pasta structures 
in Fig.\ \ref{figKEOScompare}.
We see again that the density range of the mixed phase is
narrower in the case of the ``no Coulomb screening'' calculation than
in the full calculation, while the EOS is almost the same.
A remarkable difference is the cell size,
especially near the onset density of kaonic pasta, for $\rho_B<0.5$ fm$^{-3}$.
The cell size given by the full calculation is always larger than that
given by the ``no Coulomb screening'' calculation.

\begin{figure}
\begin{minipage}{0.38\textwidth}
  \includegraphics[width=.98\textwidth]{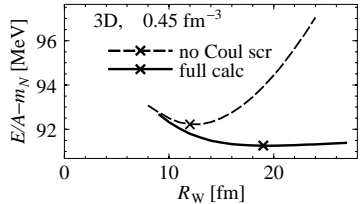}
\end{minipage}\hspace{\fill}
\begin{minipage}{0.58\textwidth}
\caption{
  The cell size $R_{\rm W}$ dependence of the energy per nucleon.
  Crosses on the curves indicate the minimum points.
}\label{figRE}
\end{minipage}
\end{figure}

To elucidate the screening effect, we depict the $R_{\rm W}$
dependence of the energy per nucleon in Fig.~\ref{figRE}.
In a general case of 3D droplet the Coulomb energy
per particle depends on the radius by its square, while the surface energy
per particle by its inverse.
Therefore the sum of the Coulomb and surface energy has a
U-shape (cf.\ ``no Coulomb screening'') and has a minimum at a certain radius.
If the Coulomb interaction is screened, the Coulomb contribution 
will be suppressed (cf.\ the full calculation) and $R_{\rm W}$ of the minimum
point gets larger.  
Since the cell radius is approximately proportional to the droplet radius for a given
baryon density,
the above argument also applies to the droplet size.

For a long time there  existed a naive view that not all the Gibbs conditions can
be satisfied in a description by the Maxwell construction (MC)
if there are two or more independent chemical components,\cite{GS99,CGS00,gle92}
because the local charge neutrality is implicitly assumed in it.
As the result of this argument,
it was suggested  that  a broad region of the 
structured mixed phase may appear in neutron stars.
However, in recent papers\cite{voskre,emaru1,maruKaon,marurev}
we have demonstrated that
if one properly includes the Coulomb interaction,
the MC practically satisfies the Gibbs conditions
and the range of the mixed phase will be limited.
The mechanism is as follows:
Due to the Coulomb interaction between charged particles,
their density distribution changes to reduce the Coulomb energy
(Coulomb screening).
Then the size of the structure gets larger and the effect of
the surface becomes small.
The local charge density gets smaller due to the screening.
Finally the feature of the MC (local charge
neutrality and no surface) is approximately achieved.


Next we consider the surface effects.
If we artificially multiply meson masses
$m_{\sigma}$, $m_{\omega}$ and $m_{\rho}$  by a factor $c_M$,
e.g.\ $c_M =1$ (realistic case), 2.5 and 5.0,
the surface tension changes.
By the use of heavy meson masses, the binding energy
of finite nuclei (for finite $A$)
approaches to that of nuclear matter indicated by a thick gray line.
This shows that the surface tension is reduced
with increase of the meson masses, cf.\ Ref.\ \refcite{MS96}.
Using the above modified meson masses, we explore the effects of
surface tension in the following.
Right panels of Figs.\ \ref{eoscompare} and \ref{figKEOScompare}
are EOS and the sizes of the cell and the lump, but now for the case of
an artificially suppressed surface tension ($c_M=5.0$).
Comparing them with the left panels,
we see that there is almost no difference in the EOS.
However, there are two differences in the case of low-density nuclear matter.
First, the density range of pasta structure is slightly broader
for weaker surface tension.
Secondly, the cell size with a normal surface tension is
larger than the case of weaker one.
It means that weaker surface tension and stronger Coulomb repulsion
cause the similar effects on the cell size since the pasta structure
is realized by the balance of the both.

In the case of kaonic pasta, the meson masses have very small effects.
The $\sigma$, $\omega$ and $\rho$ mesons have
less contribution to the surface tension of
kaonic pasta but $K$-$N$ interaction is dominant.

\section{Quark Pasta Structures in the Hadron-Quark Mixed Phase}

Many theoretical studies have shown that the hadronic
equation of state (EOS) becomes very soft
once hyperons become components of the matter. 
As a major consequence, the maximum mass of neutron stars (NS) 
predicted using the hyperonic EOS
may remain below the current observational values of about 
1.5 solar masses.\cite{pag}
Some authors have suggested that this situation might be 
remedied by considering the yet unknown three-body forces (TBF) 
among hyperons and nucleons,\cite{nishizaki} 
while other studies have shown that a quark deconfinement phase
transition in hyperonic matter renders the EOS sufficiently stiff again 
to allow NS masses consistent with current data.\cite{qmns}

We study the phase transition between hadronic and quark matter
combining a Brueckner-Hartree-Fock (BHF) EOS of hyperonic hadronic matter 
with the standard phenomenological MIT model for the quark phase.

\subsection{Treatment of hadron-quark mixed phase}

Our theoretical framework for the hadronic matter
is the nonrelativistic BHF approach 
based on microscopic
NN, NY, and YY potentials that are fitted to scattering phase shifts, 
where possible.
Nucleonic TBF are included in order to (slightly) shift
the saturation point of purely nucleonic matter to the empirical value.
With the absence of adjustable parameters,
the BHF model is a reliable and well-controlled theoretical approach
for the study of dense baryonic matter.

The basic input quantities in the Bethe-Goldstone equation
are the NN, NY, and YY potentials.
In this work we use the Argonne $V_{18}$ NN potential 
supplemented by
the Urbana UIX nucleonic TBF 
and the Nijmegen soft-core NSC89 NY potentials\cite{nsc89}
that are well adapted to the existing experimental NY scattering data
and 
$\Lambda$ hypernuclear levels.\cite{yamamoto,vprs01} 
With these potentials, the various $G$-matrices are evaluated
by solving numerically the Bethe-Goldstone equation. 
Then the total nonrelativistic hadronic energy density, $\epsilon_H$,
can be evaluated:
\begin{eqnarray}
 \epsilon_H \!&=& \!\!\!\!
 \sum_{i=n,p,\Lambda,\Sigma^-}
 \sum_{k<k_F^{(i)}}
 \left[ T_i(k) + {1\over2} U_i(k) \right] \:,
\label{e:eps}
\end{eqnarray}
with
$T_i(k) = m_i + {k^2\!/2m_i}$
and 
the various single-particle potentials 
 $U_i(k)$
are determined self-consistently from the $G$-matrices.


For the quark EOS, we use the MIT bag model with 
massless $u$ and $d$ quarks and massive $s$ quark with $m_s= 150$ MeV.
The quark matter energy density $\epsilon_Q$ 
can be expressed as a sum of the kinetic term
and the leading-order one-gluon-exchange term\cite{jaf,tama}
for the interaction energy
proportional to the QCD fine structure constant $\alpha_s$, 
and the bag constant $B$ which is the energy density difference between 
the perturbative vacuum and the true vacuum,
\begin{eqnarray}
 \epsilon_Q({\bf r}) &=& B + \sum_f \epsilon_f (\rho_f({\bf r})) \:.
\end{eqnarray}
%
%
%

\def\tsurf{\sigma}
\def\vc{V_{\rm C}}
\def\rv{{\bf r}}


The numerical procedure to determine the EOS and the
geometrical structure of the MP is similar to that 
%
explained in Secs.\ 2 and 3.
We employ a Wigner-Seitz approximation in which
the whole space is divided into equivalent Wigner-Seitz 
cells with a given geometrical symmetry.
A lump portion made of one phase is embedded in the other phase and thus 
the quark and hadron phases are separated in each cell.
A sharp boundary is assumed between the two phases and the surface energy
is taken into account in terms of a surface-tension parameter 
$\tsurf=40\;\rm MeV\!/fm^2$. 
We use the Thomas-Fermi approximation for the density profiles of
hadrons and quarks, 
while the Poisson equation for the Coulomb potential $\vc$ is explicitly solved.
The energy density of the mixed phase is thus written as
\begin{equation}
 \epsilon_M = {1\over {V_W}} \left[ 
 \int_{V_H}\!\!\!\! d^3 r \epsilon_H({\rv})+
 \int_{V_Q}\!\!\!\! d^3 r \epsilon_Q({\rv})+
 \int_{V_W}\!\!\!\! d^3 r \left( \epsilon_e({\rv}) + {(\nabla \vc({\rv}))^2\over 8\pi e^2} \right)
 + \tsurf S \right] \:,
\end{equation}
where the volume of the Wigner-Seitz cell $V_W$ is the sum of 
those of hadron and quark phases $V_H$ and $V_Q$,
and $S$ the quark-hadron interface area.
$\epsilon_e$ indicates the kinetic energy density of 
electron.
The energy densities $\epsilon_H$, $\epsilon_Q$ and $\epsilon_e$ are 
$\rv$-dependent since they are functions of
local densities $\rho_a(\rv)$ ($a=n,p,\Lambda,\Sigma^-,u,d,s,e$). 
For a given density $\rho_B$, the optimum dimensionality of the cell,
the cell size $R_W$, the lump size $R$,
and the density profile of each component
are searched for to give the minimum energy density.
For details, see Ref.\ \refcite{let}.

\subsection{Hadron-quark pasta structure}

\begin{figure*}
\begin{minipage}{0.48\textwidth}
\includegraphics[width=0.90\textwidth]{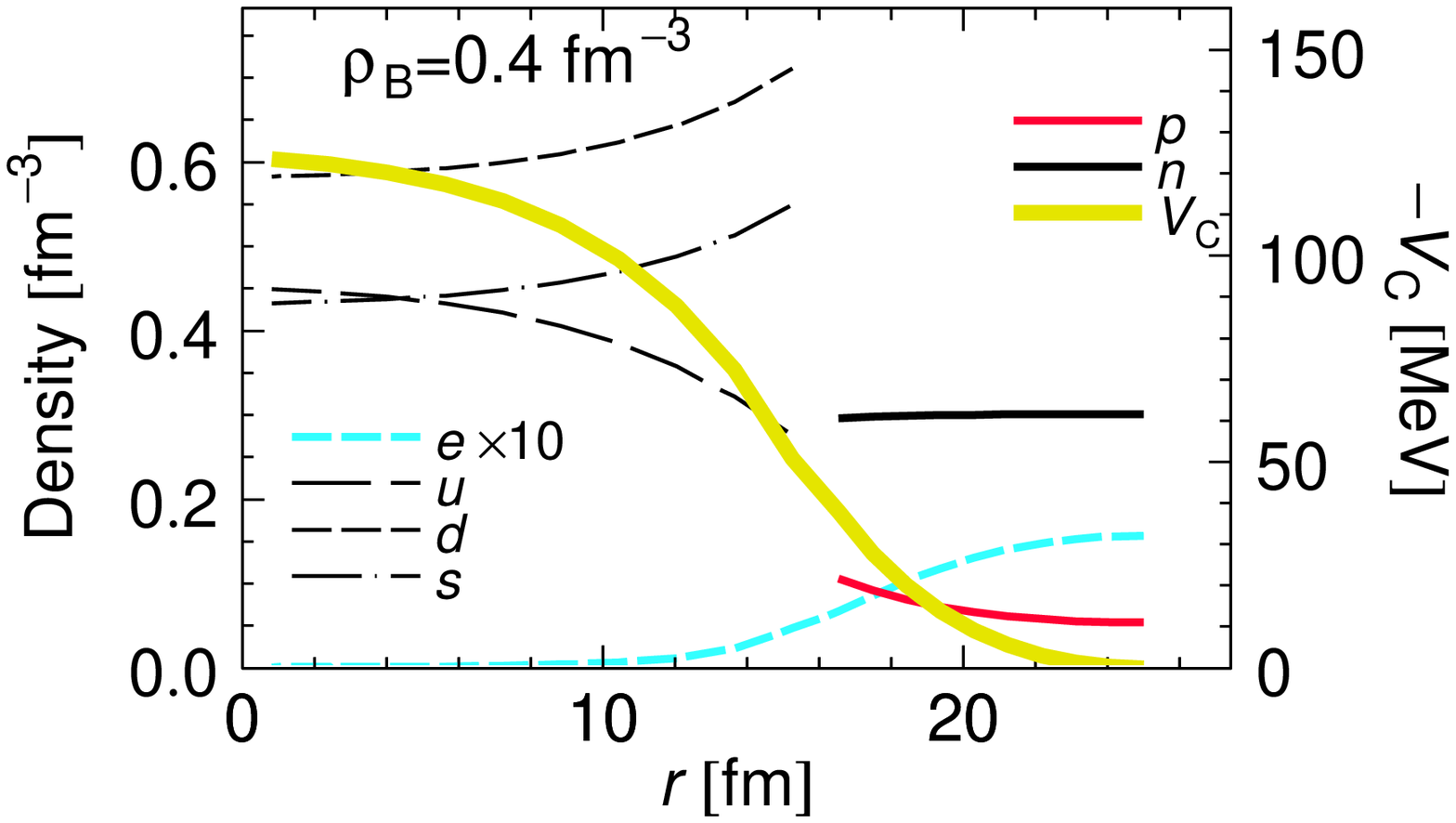}
\caption{
Density profiles 
and Coulomb potential $\vc$ 
within a 3D (quark droplet) Wigner-Seitz cell
of the MP at $\rho_B=0.4$ fm$^{-3}$.
The cell radius and the droplet radius are $R_W=26.7$ fm
and $R=17.3$ fm, respectively.
}
\label{figProf}
\end{minipage}
%
\hspace{\fill}
\begin{minipage}{0.48\textwidth}
\includegraphics[width=0.90\textwidth]{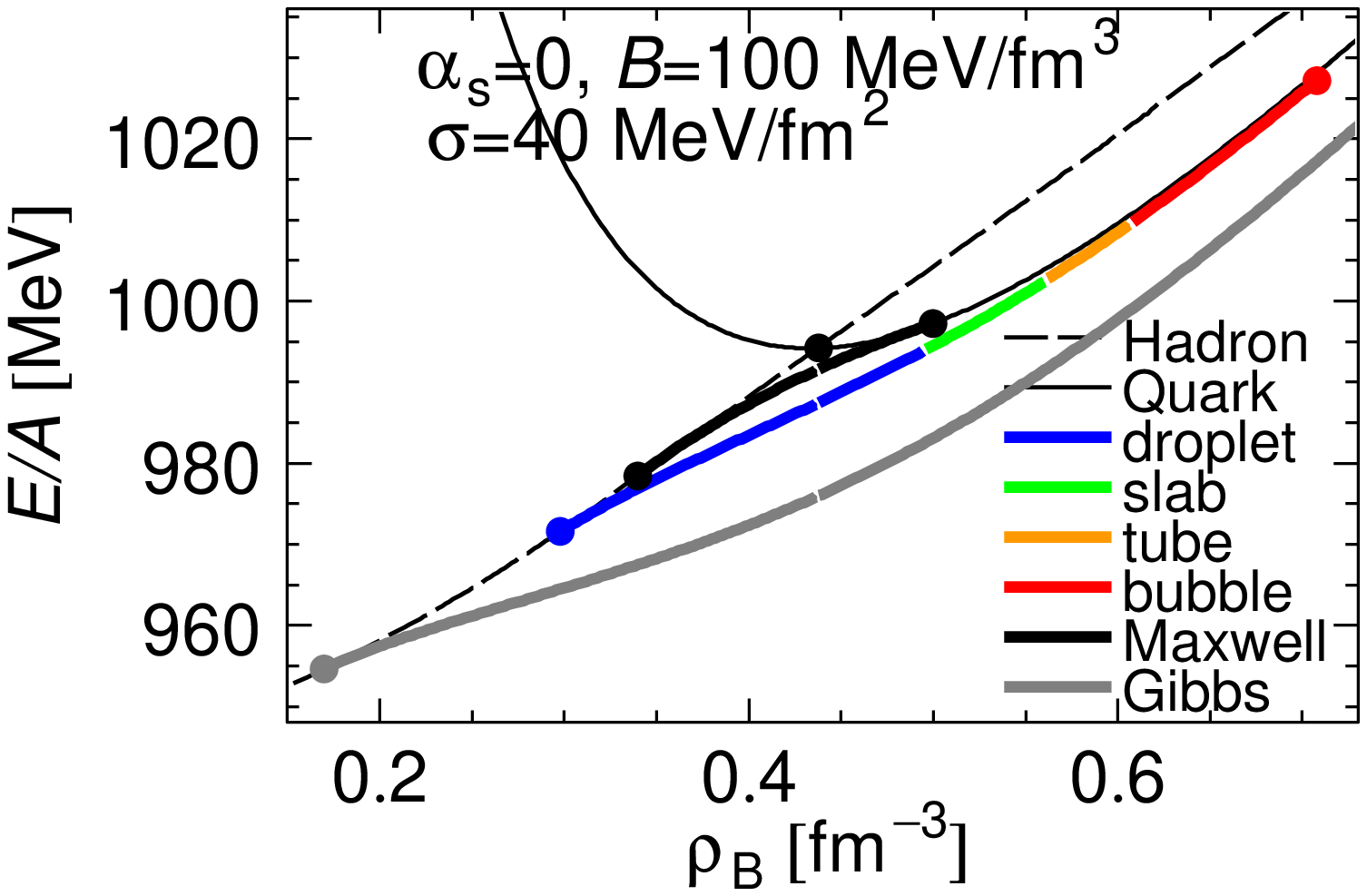}
\caption{
EOS of the MP (thick curves)
compared with pure hadron and quark phases (thin curves).
Different segments of the MP are chosen by minimizing the energy.
}
\label{figEOS}
\end{minipage}
\end{figure*}

Figure~\ref{figProf} illustrates the outcome of this procedure,
showing the density profile in a 3D cell for $\rho_B=0.4$ fm$^{-3}$.
One can see the non-uniform density distribution of each particle species
together with the finite Coulomb potential.
The quark phase is negatively charged, so that 
$d$ and $s$ quarks are repelled to the phase boundary, 
while $u$ quarks gather at the center.
The protons in the hadron phase are attracted by the negatively charged 
quark phase, while the electrons are repelled.

Figure~\ref{figEOS}  compares the resulting energy per baryon of 
the hadron-quark MP with that of the pure hadron and quark phases
over the relevant range of baryon density. 
The thick black curve indicates the case of the Maxwell construction,
while the colored line indicates the MP
in its various geometric realizations,
starting at $\rho_B=0.326$ fm$^{-3}$ with a quark droplet structure
and ending at $\rho_B=0.666$ fm$^{-3}$ with a hadron bubble structure.
The energy of the MP is only slightly lower than that of the MC, 
and the resultant EOS is similar to the MC one.

\begin{figure}
\vspace{-3mm}
\begin{minipage}{0.35\textwidth}
\includegraphics[width=0.98\textwidth]{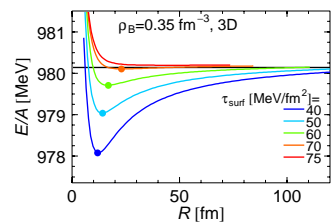}
\end{minipage}\hspace{\fill}
\begin{minipage}{0.58\textwidth}
\caption{
Droplet radius ($R$) dependence of the energy per baryon
for fixed baryon density $\rho_B=0.35$ fm$^{-3}$
and different surface tensions.
The quark volume fraction $(R/R_W)^3$ is fixed for each curve.
Dots on the curves show the local energy minima.
The black line shows the energy of the MC case.
}
\label{figRdep}
\end{minipage}
\vspace{-3mm}
\end{figure}

If one uses a smaller surface tension parameter $\tsurf$, 
the energy gets lower and the density range of the MP gets wider.
The limit of $\tsurf=0$ leads to a bulk application of 
the Gibbs conditions without the Coulomb and surface effects.\cite{let}  
On the other hand, using a larger value of $\tsurf$, 
the geometrical structures increase in size and
the EOS gets closer to that of the MC case. 
Above a limiting value of $\tsurf \approx 65\;\rm MeV\!/fm^2$ 
the structure of the MP becomes mechanically unstable\cite{vos}: 
for a fixed  volume fraction $(R/R_W)^3$
the optimal values of $R$ and $R_W$ diverge
and local charge neutrality is recovered in the MP, 
where the energy density equals that of the MC 
(see Fig.~\ref{figRdep}).

\section{Summary}

We have discussed the appearance and properties of the pasta structures and the EOS of
matter in the context of the first-order phase transition
in the cases of (I) low density nucleon matter, (II) kaon condensation and 
(III) hadron-quark deconfinement transition.
We have shown a general feature that the Coulomb screening and the stronger 
surface tension enlarge the structure size, and limit the density region
of the mixed phase.
If the surface tension is strong enough, the Maxwell construction (MC) becomes effective.
Actually, we have seen that the EOS of the mixed phase is close to that of the MC.

For each case, we have shown the following:
The Coulomb screening effects in (II) and (III) are stronger than in (I)
due to the higher charge density.
The surface tension is dominated by $K$-$N$ interaction for (II).
Though we have not reported here, the bulk property of
a compact star is strongly affected by (III).\cite{marufull}

\section*{Acknowledgments}

This work has been done by the collaboration with
T.~Tanigawa, D.~N.~Voskresensky T.~Endo and H.-J.~Schulze.

\end{document}